\documentclass[a4paper]{article}
\usepackage{qcircuit}
\usepackage{graphicx}
\usepackage{subcaption}
\usepackage{amsmath}
\usepackage{pythonhighlight}
\usepackage{listings}
\usepackage{helvet} % for PDFLaTeX
\usepackage[utf8x]{inputenc}
\usepackage[T1]{fontenc}
\usepackage{authblk}
\usepackage{verbatim}
\usepackage[a4paper,top=3cm,bottom=3cm,left=2cm,right=3cm,marginparwidth=1.75cm]{geometry}
\input{amssym}

\usepackage[colorinlistoftodos]{todonotes}
\usepackage[colorlinks=true, allcolors=blue]{hyperref}
\usepackage{authblk}

\usepackage{setspace}

\title{\bf AriaQuanta: A Quantum Software for Quantum Computing  }

\author[1]{ A. Cheraghi }
\author[2]{H.Davoodi Yeganeh\footnote{corresponding author: h.yeganeh@ut.ac.ir, h.yeganeh@ariaquanta.com}}
\affil[1,2]{Quantum Research Center, Shahid Sattari University of Aeronautical Sciences and Technology, Tehran, Iran}

\affil[1,2]{AriaQuanta Quantum Computing Center, Tehran, Iran}

\date{}

\begin{document}

\maketitle

\begin{abstract}
We introduce AriaQuanta, a powerful and flexible tool for designing, simulating, and implementing quantum circuits. This open-source software is designed to make it easy for users of all experience levels to learn and use quantum computing. The first version includes a compiler for implementing various quantum circuits and algorithms. Additionally, parametric circuits allow for the implementation of variational quantum algorithms, and various noise models are available for simulating noisy circuits. We performed numerous numerical simulations on AriaQuanta in various applications, including quantum algorithms and noisy circuits. The results, compared with popular counterparts, demonstrate the high performance of AriaQuanta.

\end{abstract}

\section{Introduction}
\setstretch{1.4}
Quantum computers, utilizing the principles of quantum physics, can solve problems that classical computers cannot. In recent years, experimental quantum computing has made remarkable progress \cite{r111,r222}, demonstrating quantum supremacy and quantum advantages \cite{r333,r444,r555,r666}.
The applications of quantum computers span a wide range of fields, including mathematics, machine learning, medicine, finance, and economics\cite{r700,r800,r900}. Consequently, they have garnered significant attention over the past decade \cite{r1,r10}.
To solve a problem using quantum computing, the problem is first defined, followed by the design of a quantum circuit and algorithm to address it. Quantum hardware or a quantum simulator can be used to run and test a quantum algorithm, necessitating the availability of quantum software (quantum compiler). Currently, quantum hardware is accessible via the cloud and can be utilized with a software stack. Quantum compilers generally convert high-level languages into hardware instructions to execute a quantum algorithm \cite{r7,r8}.
Recently, the development of quantum software tools has accelerated, resulting in numerous quantum software packages available from various platforms and sources. Popular quantum software includes Cirq \cite{r4}, Qiskit \cite{r5}, ProjectQ \cite{r6}, and PennyLane \cite{r6666}. For a comprehensive review of quantum software, see refs \cite{r2,r3,r9}. Most quantum programming tools have been developed for research purposes in simulating quantum circuits and algorithms. Although a few of these tools have been used for real implementations on quantum hardware, the majority remain focused on theoretical applications.
Here, we introduce an open-source quantum software platform designed to simulate quantum circuits and algorithms. AriaQuanta is a quantum programming language implemented in Python, specifically created for the analysis, construction, and execution of quantum programs. It can interface with various quantum hardware and functions as a simulator, resource estimator, and emulator, allowing users to test, run, and debug quantum algorithms.
Additionally, the AriaQuanta compiler supports GPU usage, enabling faster program execution.The Ariaquanta emulator can also run a program serially or in parallel (single-core and multi-core) on the CPU. Since this platform is written in Python, it is well-suited for Noisy Intermediate Scale Quantum (NISQ) devices. AriaQuanta offers multiple backends for simulating quantum algorithms and can run on GPUs, which allows it to execute programs more quickly than similar platforms. Compared to other quantum programming platforms, AriaQuanta encompasses all their features while also being GPU-capable and run a program serially or in parallel . The framework will be fully introduced and reviewed in the following sections.

\section{The AriaQuanta Architecture}\label{sec1}

\subsection{Simplified Architecture}
\begin{figure}[h!]
\centering
\includegraphics[width=9.5cm]{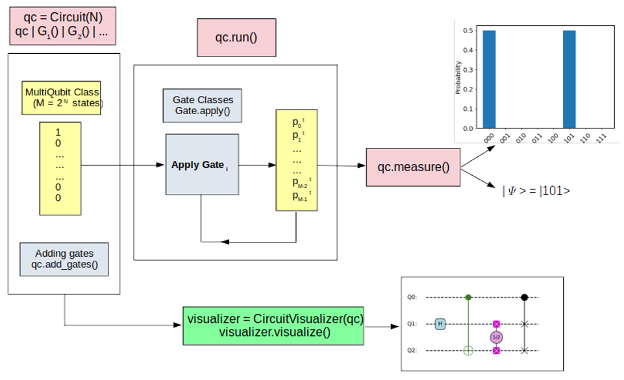}
\caption{General scheme of the modified version. First, we define the quantum circuit $qc$ with $N$ qubit input and then add the required gates to it. In the $qc.run$ section, the gates are applied to the system state vector in order and finally $qc.measure$ performs the measurement. The visualizer class is also used to draw the circuit. }\label{fig1}
\end{figure}

The AriaQuanta platform is a Python-based library for quantum computing. The first step in implementing quantum computing involves defining qubits, gate operators, and circuits, and then applying these operators to the N-qubit states. In its initial implementation, AriaQuanta can simulate quantum circuits with any number of qubits and gates, as well as perform quantum measurements on the state vector. Figure \ref{fig1} shows a simplified schematic of the platform structure. This figure illustrates the overall operation of the quantum circuit, including the measurement process and the circuit diagram. This structure facilitates the execution of quantum programs, making it user-friendly.
It is important to note that in this simplified version, the simulation operation is performed within the same circuit class, eliminating the need to use the backend. In this case, the user can execute a quantum circuit with simple commands. As shown in Figure 1, the process begins with defining the qubits, followed by adding various types of gates—single-qubit, two-qubit, three-qubit, control gates  and arbitrary quantum gates—to the circuit. Finally, the measurement is performed, and the output is provided as a state vector or probability.

All commonly used gates are defined within the platform and are easy to use. Additionally, arbitrary gates can be easily defined. In the circuit drawing section, different gates are displayed in various colors for easier identification and visual appeal. Single-qubit gates are shown in blue, two-qubit gates in purple, and three-qubit gates in black. All circuit information is available, including the number of qubits, gates, size, and depth. 
%We introduce the main features of AriaQuant in modified version, starting with a simple example in Listing 1.
\begin{comment}
\begin{python}
# AriaQuanta code for a Bell state

from AriaQuanta import*
import numpy as np

1 qc = Circuit(2)    # Defining circuit with 2 qubit
2 qc | H(0) | CX(0,1) 
3 visualizer = CircuitVisualizer(qc)   # Visualization
4 fig, ax = visualizer.visualize()
5 result = qc.run() # Run the circuit and output is a state vector
6 measurement, measurement_index, probabillities = qc.meaure(). # Measurment

\end{python}

In line 1, we define a circuit with two qubits. In line 2, Hadamard and CNOT gates are added to the circuit. Lines 3 and 4 provide the visualization of the quantum circuit. Finally, in line 5, the output of the circuit is plotted as a state vector, and in line 6, as a probability. 
%The output of the code is as follows:
%\begin{figure}[h!]
%\centering
%\includegraphics[width=11cm]{P2}
%\caption{Output AriaQuanta code for a Bell state }\label{fig2}
%\end{figure}
%\newpage

\end{comment}

\subsection{General  Architecture}

\begin{figure}[h!]
\centering
\includegraphics[width=11cm]{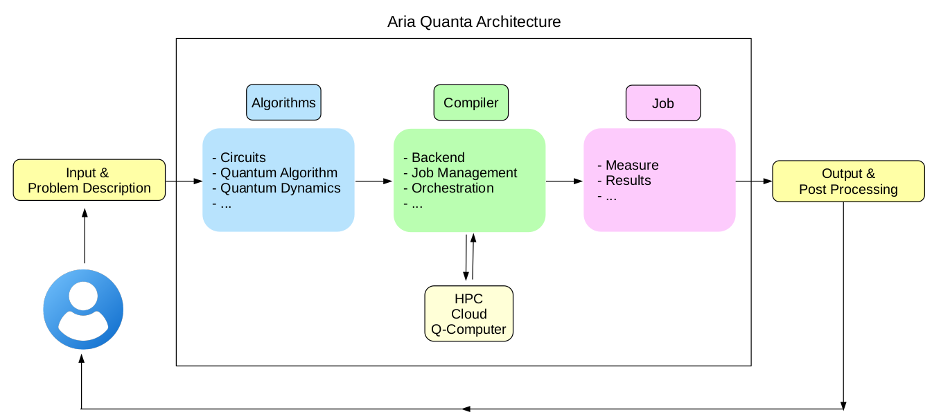}
\caption{General scheme of AriaQuanta  }\label{fig3}
\end{figure}

Before delving into the various functions of the Aria Quanta, we will first provide an overview of its general architecture. A simplified version of this architecture was presented in section \ref{sec1} . In this section, we expand on that framework and develop it further to include additional applications. Figure \ref{fig3} illustrates the general architecture of AriaQuanta.

This architecture includes features such as the ability to integrate various functions, communicate with external computing systems, manage computing tasks efficiently, and provide a user-friendly interface. In the general architecture, a quantum circuit or algorithm is first defined, similar to the previous section. It is then sent to a compiler that integrates a backend and task management and after executing the program, the results are obtained. It is important to note that the program can be executed using a CPU, QPU, or HPC. The results include the state vector, density matrix, and probability distribution of states. Each part will be explained in detail in the following sections.

\section{Overview of AriaQuanta}

In this section, we provide a detailed overview of AriaQuanta’s functionalities, including the simulation of  quantum circuits, as well as handling their noisy counterparts.

\subsection{Qubit initialization}

To implement a quantum circuit, we first need to define the qubits. AriaQuanta offers two types of qubits: single-qubit and multi-qubit. Both types include the initial state vector. The single-qubit class creates a qubit object, which is then used to define a multi-qubit class. The multi-qubit class contains 5 qubits and includes properties such as the number of qubits, the state vector, and a list of all qubits. Listing 1 shows the example code for initializing  quantum states with 3 qubits.

\begin{python}
Code1. Qubit initialization

from AriaQuanta.aqc.qubit import Qubit, MultiQubit
onequbit = Qubit('q0')  #Defining 1 qubit using Qubit
qubits = MultiQubit(3)  #Defining 3 qubits using MultiQubit
print("\nstate is:", onequbit.state)
print("\nname is:", onequbit.name)
print('\nstate of qubit[0] is:', onequbit_multi.qubits[0].state)
print('\nname of qubit[0] is:', onequbit_multi.qubits[0].name)
\end{python}

\subsection{Quantum gates}

The subsequent step in quantum circuit implementation involves the utilization of quantum gates.. In general, a quantum gate is a unitary operator ($UU^\dagger=I$), divided into three main categories based on the number of inputs and outputs: single-qubit, two-qubit, and three-qubit gates. Multi-qubit gates are constructed using these basic gates. AriaQuanta consists of five classes of gates: single-qubit gates, two-qubit gates, three-qubit gates, control gates, and arbitrary gates. Note that the control gate can be categorized as a two-qubit or three-qubit gate, but it is separated here for simplicity. The arbitrary gate can also be defined by its matrix. Currently, single-qubit gates include  
 {\fontfamily{pcr}\selectfont I,X,Y,Z,S,H,P,T,Xsqrt,GlobalPhase,Rot,RX,RY,RZ} gates.
 
\noindent Two-qubit gates include {\fontfamily{pcr}\selectfont SWAP,ISWAP,SWAPsqrt,SWAPalpha,Magic,RXX,RYY,RZZ,RXY,

Barenco,Berkeley,Canonical,Givens} gates. 
\noindent Three-qubit gates include  {\fontfamily{pcr}\selectfont Toffoli,Margolus,Fredkin} gates. 
Control gates gates include  {\fontfamily{pcr}\selectfont CX,CZ,CP,CU(arbitrary  U),CS,CSX(Controlled-Sqrt-X)} gates.

The general utilization of  gate is {\fontfamily{pcr}\selectfont G(i)},  {\fontfamily{pcr}\selectfont G(i,j)} and {\fontfamily{pcr}\selectfont G(i,j,k)} for single-qubit gates, two-qubit gates and three-qubit gates respectively and $i$, $j$, $k$ represent the qubits. If the gate is parametric such as {\fontfamily{pcr}\selectfont RX($\theta$)}, in addition to the qubit, the parameter value is also given i.e {\fontfamily{pcr}\selectfont G(i, parameter)} or {\fontfamily{pcr}\selectfont G(i,j, parameter)} where the number of parameters can be different for gates.
Parametric gates are essential for variational quantum  circuits and algorithms in NISQ area. Aria Quanta provides t single-qubit and two-qubit parametric gates: $R_x(\theta)$, $R_y(\theta)$, $R_z(\theta)$, $U(\theta,\phi,\gamma)$ and $R_{xx}(\phi)$, $R_{yy}(\phi)$, $R_{zz}(\phi)$, $R_{xy}(\phi)$, $Barenco(\alpha,\phi,\theta)$, $Canonical(a,b,c)$, $Givens(\theta)$. In addition, a parametric control gate can be defined in arbitrary control gate $CU$. 
Listing 2 shows the example code for using quantum gates

\begin{python}
Code2. Implement quantum gates

from AriaQuanta.aqc.gatelibrary import*
import numpy as np

rxxgate = RXX(np.pi)
print("\nname is:", rxxgate.name)
print("\nmatrix is:", rxxgate.matrix)
print("\nqubits are:", rxxgate.qubits)

n=0,m=1
Z(n)   #apply gate Z on qubit 0
CX(n,m) #apply CNOT gate, n is control qubit and m is target qubit
RX(theta, m)  #apply RX gate on qubit 1

\end{python}

In addition to standard and custom gates, noise in quantum operations is described through three main types: phase flip, bit flip, and depolarization. AriaQuanta utilizes these noise models to implement noisy quantum circuits. These noises can be used in a quantum circuit with the {\fontfamily{pcr}\selectfont BitFlipNoise(p,n)}, {\fontfamily{pcr}\selectfont DepolarizingNoise(p,n)}, {\fontfamily{pcr}\selectfont PhaseFlipNoise(p,n)}, where {\fontfamily{pcr}\selectfont p} is the probability of the noise effect and  {\fontfamily{pcr}\selectfont n}  is the number of qubits.

\subsection{Quantum circuits}
A quantum circuit represents qubits and quantum gates. In AriaQuanta, circuits can be easily defined and manipulated by calling the {\fontfamily{pcr}\selectfont Circuit
} and adding gates. There are two methods to add gates to a circuit: the {\fontfamily{pcr}\selectfont add\_gate} method and the  {\fontfamily{pcr}\selectfont \_or\_}  operator i.e   {\fontfamily{pcr}\selectfont  |} operator. The circuit methods include circuit depth, circuit size, circuit size, number of qubits, and initial state vector. Also, with the method  {\fontfamily{pcr}\selectfont to\_gate} a quantum circuit can be converted into a gate and used in other circuits and the number of classical bits that can be designated for storing the state after measurement. 
To draw a quantum circuit, the class
{\fontfamily{pcr}\selectfont CircuitVisualizer} and method {\fontfamily{pcr}\selectfont visaulize} are used. In Ariaquanta, different colors are assigned to different gate categories.
 Listing 3 shows the example code for  employ quantum circuits

\begin{python}
Code3. implementation of quantum circuits

from AriaQuanta.aqc.circuit import Circuit
from AriaQuanta.aqc.gatelibrary import*
from AriaQuanta.aqc.visualization import CircuitVisualizer

qc = Circuit(4)   # circuit white 4 qubits
qc | X(0)|X(1)
qc | CX(0,2) | CX(1,2)
qc | CCX(0,1,3)
visualizer = CircuitVisualizer(qc)
visualizer.visualize()

print("\nnumber of qubits are: ", qc.num_of_qubits)
print("\nnumber of classical bits are: ", qc.num_of_clbits)
print("\ncounting all the gates: ", qc.gatesinfo)
print("\nwidth of the circuit: ", qc.width)
print("\nsize of the circuit: ", qc.size)
print("\ndepth of the circuit: ", qc.depth)
print("\ninitial state is: ", qc.initial_state)
print("\nstate vector before running is the initial state: ", qc.statevector)
print("\ndensity matrix before running, the matrix represenation of the initial state: ", qc.density_matrix)

\end{python}

\subsection{Running quantum circuits}

Once the qubits are prepared and the quantum gates are added to the circuit, the circuit must be executed, and the results displayed. In AriaQuanta, there are two ways to run the circuit. As discussed in the section \ref{sec1}, we consider two architectures: simplified and general. In the simplified architecture, the circuit is executed using the command  {\fontfamily{pcr}\selectfont qc.run()}. After running the circuit, we can obtain the following outputs using the command {\fontfamily{pcr}\selectfont qc.measure()}:  (i) The probability and probability diagram of the states, (ii) The final state of the system, and (iii) The index of the final state.  Listing 4 shows the example code for  running quantum circuits in simplified approach.

\begin{python}
Code4. Running quantum circuits 

from AriaQuanta.aqc.gatelibrary import H, X
from AriaQuanta.aqc.circuit import Circuit, sv_to_probabilty
import numpy as np

qc = Circuit(2)
qc |  H(0) | CX(0,1)
qc.run()
measurement = qc.measure_all()
print("\nmeasurement result: ", measurement)
print("\nstate vector: ", qc.statevector)
print("\ndensity matrix: ", qc.density_matrix)
probability = sv_to_probabilty(qc.statevector, plot=True)
print("\nprobability: ", probability)
\end{python}
Generally, circuit running is performed using the backend. In this case, a circuit can be executed multiple times in parallel or in series. Additionally, we can specify whether the code is executed on the CPU or GPU or QPU. The output can be a state vector, a density matrix, or a probability distribution of states. In Ariaquanta, these tasks are handled by the simulator, which is defined in the backend.  Note that any desired qubit can be measured in this case (with classical bit), and the number of circuit executions can be divided arbitrarily. In other words, it can be specified how many executions are run in parallel for each of the n executions in each step. Listing 5 shows the example code for  running quantum circuits in general approach. 

\begin{python}
Code5. Running quantum circuits in general approach

from AriaQuanta.aqc.gatelibrary import *
from AriaQuanta.aqc.circuit import Circuit
from AriaQuanta.aqc.measure import MeasureQubit
from AriaQuanta.backend.simulator import Simulator
from AriaQuanta.backend.result import plot_histogram
from AriaQuanta.aqc.noise import BitFlipNoise

qc = Circuit(3)
qc |  H(0) | CX(0,1) | H(2) | BitFlipNoise(0.1,0) | MeasureQubit([0,1])
sim = Simulator()
result = sim.simulate(qc, 1000, 1)
counts, probability = result.count()
print('\ncounting measurement on the result:\n', counts)
print('\nprobability of each state:\n', probability)
plot_histogram(counts)
plot_histogram(probability)

\end{python}

\subsection{ Quantum algorithm}

Quantum algorithms refer to methods that utilize the principles of quantum computing to solve complex mathematical problems more efficiently. These algorithms leverage physical properties such as quantum entanglemet and superposition to perform calculations. Currently, the Grover\cite{g1}, Shor\cite{s1}, Deutsch-Jozsa\cite{dj}, and Bernstein–Vazirani\cite{b1} algorithms and Phase estimation, Fourier transform, and inverse Fourier protocols have also been defined. have been implemented in AriaQuanta. Additionally, the  Variational Quantum Eigensolver(VQE)\cite{vqe}, and Quantum Approximate Optimization Algorithm(QAOA)\cite{qaoa} algorithms have been implemented for use in NISQ devices. The number of available algorithms is expected to increase in future versions. Furthermore, based on the definition of noise, the effects of noise can be observed in these algorithms.
To use them, simply call the algorithm class as {\fontfamily{pcr}\selectfont from AriaQuanta.algorithms import}.  Listing 6 shows the example code for  implement Deutsch-Jozsa algorithm.

\begin{python}
Code6. Deutsch Jozsa Algorithm

from AriaQuanta.aqc.circuit import Circuit
from AriaQuanta.algorithms import dj
from AriaQuanta.aqc.visualization import CircuitVisualizer
from AriaQuanta.aqc.circuit import Circui

1 n_qubits = 4
2 qc_dj = dj(n_qubits, is_constant=False) # number of qubits and constant/balanced
3 qc_dj.run()
4 measurement, measurement_index, probabilities = qc_dj.measure()
# Analyze the result
5 measurement_n_qubits = measurement[1:n_qubits+1]
6 if measurement_n_qubits == '0' * n_qubits:
7    output = "measurement = " + measurement_n_qubits + ", Function is constant"
8 else:
         output = "measurement = " + measurement_n_qubits + ", Function is balanced"

\end{python}

\section{AriaQuanta's performance}
\begin{figure}[h]
    \centering
    \begin{subfigure}[b]{0.49\textwidth}
        \includegraphics[width=\textwidth]{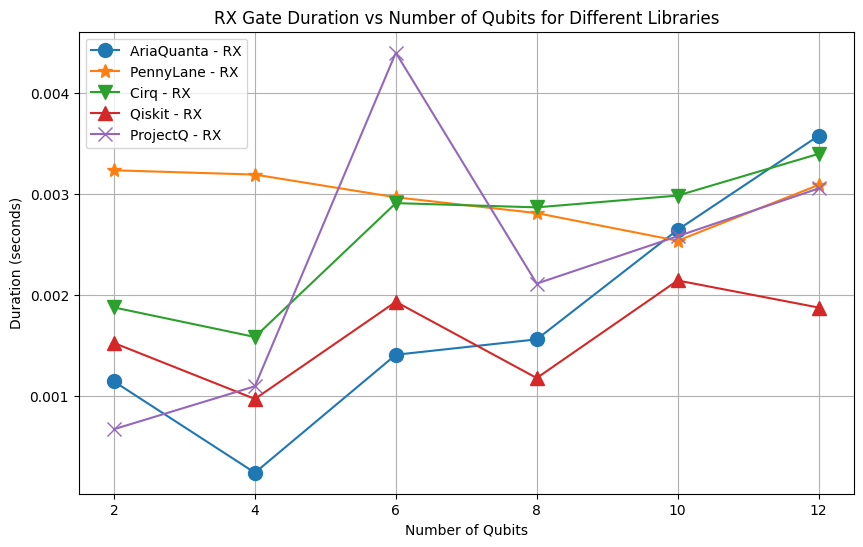}
        \caption{Comparison of runtime performance for $R_x$}
        \label{fig:image1}
    \end{subfigure}
    \hfill
    \begin{subfigure}[b]{0.49\textwidth}
        \includegraphics[width=\textwidth]{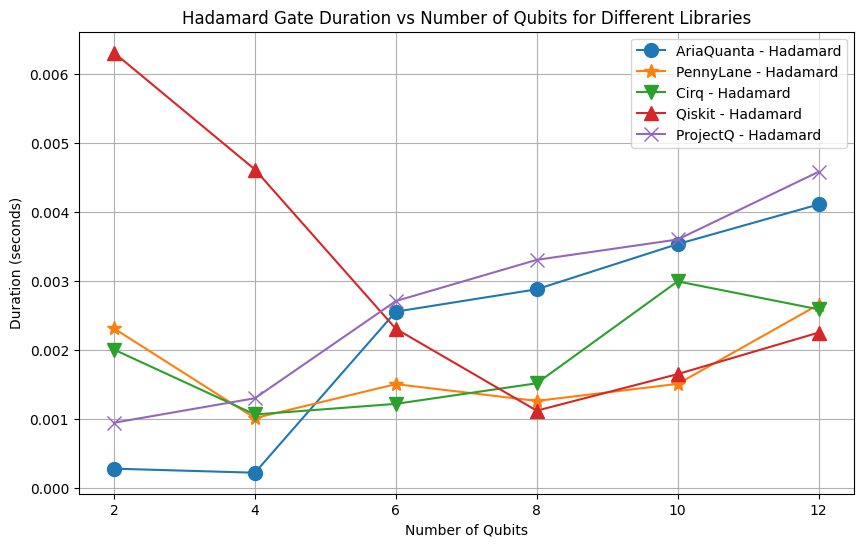}
        \caption{Comparison of runtime performance for H}
        \label{fig:image2}
    \end{subfigure}
    \hfill
    \begin{subfigure}[b]{0.49\textwidth}
        \includegraphics[width=\textwidth]{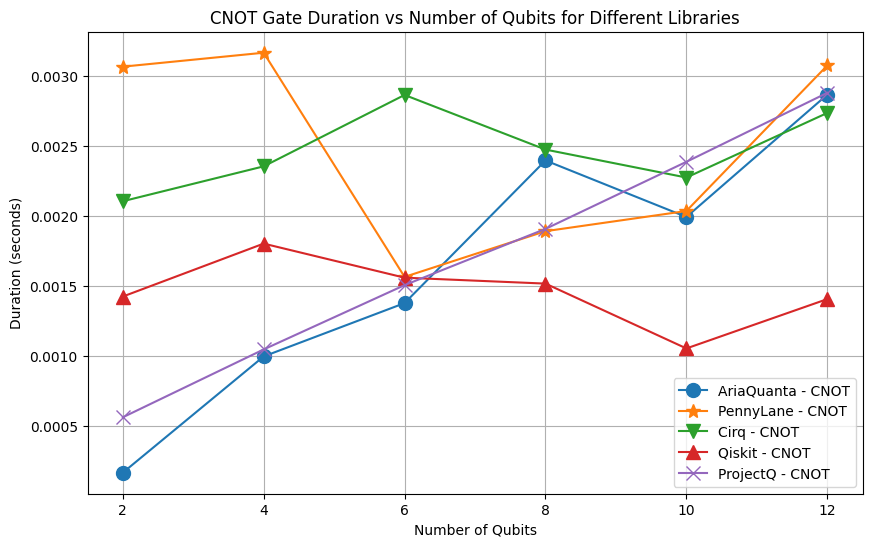}
        \caption{Comparison of runtime performance for CNOT}
        \label{fig:image3}
    \end{subfigure}
    \caption{A comparison of  runtime performance}
    \label{fig4}
\end{figure}

This section evaluates the performance of Ariaquanta and compares it with other popular quantum simulators: Qiskit, Cirq, ProjectQ, and Pennylane. Ariaquanta's capabilities are assessed based on several criteria, including execution speed, accuracy, and the ability to handle different quantum algorithms. Our analysis includes benchmarking tests where the same set of quantum gates, variational quantum circuit, noisy circuit and algorithms are executed on each simulator. For performance comparison, we used a device equipped with Intel(R) Corei7-7500U CPU, running Ubuntu 22.04.05.  To evaluate the performance of AriaQuanta, we conducted tests on the implementation of RX, Hadamard, and CNOT gates across a range of 1 to 12 qubits. The results, illustrated in Figure \ref{fig4}, reveal that the execution time of the circuit increases exponentially as the number of qubits rises, which is consistent with our expectations. As observed, AriaQuanta demonstrates superior performance when compared to other quantum software.

\section{Summary}
 We have developed an open-source quantum software, AriaQuanta, designed for simulating quantum circuits and algorithms. It is user-friendly and accessible for all researchers in the field of quantum computing. AriaQuanta supports the execution of conventional and variational quantum circuits and algorithms, as well as noisy circuits. Additionally, this software demonstrates superior performance compared to other quantum software available.
 \section*{Funding Declaration}
 This research did not receive any specific grant from funding agencies in the public, commercial, or not-for-profit sectors.

\bibliographystyle{unsrt}
\bibliography{vosq-1}

\newpage

\section*{Examples}

This section presents several examples that have been implemented in Ariaquanta.

\subsection*{E1. Quantum Teleportation}

The implementation of quantum teleportation can be outlined as follows:
\begin{python}

from AriaQuanta.aqc.circuit import Circuit
from AriaQuanta.aqc.gatelibrary import H, CX, Z, X
from AriaQuanta.aqc.measure import MeasureQubit
from AriaQuanta.aqc.operations import If_cbit
from AriaQuanta.aqc.qubit import create_state

# define qubits with optional states

alpha = 0.7
# create qubit=0 with state alpha|0>+beta|1>
q0 = create_state(0,alpha) 

# create qubit=1 with state |0>
q1 = create_state(1,1)

# create qubit=2 with state |0>
q2 = create_state(2,1)

# add qubits to a circuit
qc = Circuit(3, list_of_qubits=[q0, q1, q2])

# Make the shared entangled state 
qc | H(1)
qc | CX(1, 2)

# Alice applies teleportation gates (or projects to Bell basis)
qc | CX(0, 1)
qc | H(0)

# Alice measures her qubits
qc | MeasureQubit([0,1], ['a','b'])

# Bob applies certain gates based on the outcome of Alice's measurements
qc | If_cbit(['a',1], Z(2))
qc | If_cbit(['b',1], X(2))

#--------------------------------------------------
# simple run and output the statevector
# Bob checks the state of the teleported qubit
qc.run()
print("\nBob's statevector:\n", qc.statevector)

\end{python}

\subsection*{E2. Bernestein-Vazirani Algorithm}
\begin{python}
from AriaQuanta.aqc.circuit import Circuit
from AriaQuanta.aqc.gatelibrary import H, CX, Z, I
from AriaQuanta.backend.simulator import Simulator
from AriaQuanta.aqc.measure import MeasureQubit
from AriaQuanta.backend.result import plot_histogram

# Bernestein-Vazirani Algorithm for string '10' 
qc = Circuit(3)
qc | H(0) | H(1) | H(2) | Z(2) | CX(0,2) | H(0) | H(1) | MeasureQubit([0,1])
sim = Simulator()
result = sim.simulate(qc, 1000, 1)
counts, probability = result.count()
print('\ncounting measurement on the result:\n', counts)
plot_histogram(counts)
\end{python}

\section*{Documents}
For additional details, refer to the GitHub link: https://github.com/AriaQuanta/AriaQuanta-tutorial

\end{document}